\documentstyle[preprint,revtex]{aps}
\begin{document}
\draft
\begin{title}
The Loss of Unitarity in the Vicinity of a Time Machine
\end{title}
\author{Dalia S. Goldwirth}
\begin{instit}
Center for Astrophysics,  60 Garden Street, Cambridge, MA 02138,
USA
\end{instit}
\author{Malcolm J. Perry}
\begin{instit}
DAMTP, University of Cambridge, Silver Street,
Cambridge, CB3 9EW, England
\end{instit}
\author{Tsvi Piran}
\begin{instit}
Center for Astrophysics,  60 Garden Street, Cambridge, MA 02138,
USA
\end{instit}
\begin{abstract}
We construct the propagator of a non-relativistic non-interacting
particle in a  flat spacetime in which two regions have been
identified. This corresponds to the simplest ``time machine". We show
that while completeness is lost in the vicinity of the time machine
it holds before the time machine appears and it is recovered
afterwards. Unitarity, however, is not satisfied anywhere. We discuss
the implications of these results and their relationship to the loss
of unitarity in black hole evaporation.
\end{abstract}
\pacs{Ms number LV4936. PACS numbers: 04.20.Cv, 04.20.Jb, 04.60.tn}

\narrowtext

Spacetime, and the phenomenon of gravitation, are described very well at a
classical level by the theory of General Relativity.  Locally, spacetime is
isomorphic to Minkowski and there is a well defined lightcone and
microscopic Causality.  Globally however, things may be quite different.
There is nothing in the laws of classical general relativity that prevents
spacetimes from having closed causal (timelike or null) curves, that is
future directed curves through a point $p$ such that if one travels along
them towards the future, one returns to the same spacetime point.  It is
easy to find examples of spacetimes in which closed timelike curves have
always existed \cite{r1}.  None of these examples, generally referred to as
eternal time machines, look very much like our Universe. In each of these
cases, it is not possible to pose the Cauchy problem for matter fields
propagating in these spacetimes \cite{r5}, and one can therefore believe
that these spacetimes are rather pathological.

Another type of causality violation is one in which closed timelike curves
develop during the evolution of a spacetime from some reasonable initial
conditions. An example of such behavior is found in the Kerr solution which
is believed to be the endpoint of gravitational collapse with rotation. The
region in which causality violation occurs is close to the singularity and
interior to the inner horizon.  It might be the case that the Kerr example
is generic under certain circumstances, as Tipler \cite{r7} has shown that
if matter obeys the weak energy condition, and closed timelike lines
develop to the future of some Cauchy surface, then the spacetime must be
geodesically incomplete. If one believes in the cosmic censorship
hypothesis \cite{r8}, then under such circumstances, the singularity is
always enclosed by a horizon, and we conjecture that if the weak energy
condition is satisfied the closed timelike lines will also only occur in
the interior of the horizon and the physics exterior to any horizon would
always be unaffected by the paradoxes and difficulties associates with
closed timelike curves. Hawking \cite{r9} has proposed the Chronology
Protection Conjecture, presently still unproven, that would prevent
causality violation under a wide range of circumstances.

Systems that obey the weak energy conditions classically, for example
a free scalar field, do not necessarily obey it after quantization
\cite{r10,r11}.
Under these circumstances, it appears to be possible to create a region of
spacetime that includes closed timelike curves without the occurrence of
spacetime singularities, other than that associated with the chronology
horizon.  Similarly, since the laws of physics are time reversal invariant,
we expect that such regions could disappear.  This type of spacetime we
refer to as a ``time machine." Morris, Thorne and Yurtsever \cite{r12} have
shown one way that such spacetimes can arise, when a wormhole connects two
spacelike separated points in Minkowski space.  There are number of
undesirable paradoxes that arise in such a spacetime.  Recently several
authors \cite{r13,r14}, discussed the resolution of these paradoxes within
the realm of classical physics.  However, in the presence of time machines
the Cauchy problem fails to be well-posed in a very explicit way
\cite{r13}.  For each classical initial value problem, there exist an
infinite number of consistent (i.e. non paradoxical) classical evolutions.
In other words, although the paradoxes can be avoided, predictability will
still be violated.

Our aim is to explore the nature of these quantum mechanical processes in
the presence of time machines.  In the absence of any microscopic quantum
theory of gravity, we can only study quantum mechanical processes on a
fixed spacetime background.  Conceptually it is easiest to work in the
Schr\"odinger picture. Then the state at time t, $|\psi (t)\rangle $, is
determined in terms of the Hamiltonian operator, $H(t)$ and an initial
state $|\psi (0)\rangle $. We cannot use this method here because the
Hamiltonian only exists in spacetimes that are globally hyperbolic.  Any
spacetime that has closed timelike curves fails to be globally hyperbolic.

An alternative approach is to use the Feynman path integral  to find the
transition amplitude $ \langle \psi(t) | \psi (0) \rangle $.  The path
integral can be derived from the Schr\"odinger formulation for certain
classes of Hamiltonian, provided that the Hamiltonian exists \cite{r15}.
However, in the absence of a Hamiltonian, the path integral is the only
tool that there is, and we regards it as the fundamental definition.

We study a free non-relativistic particle. At a time $t_i$, the particle is
in an eigenstate of position at $x_i$, so it is in the state $|i, t_i
\rangle $. The propagator is the amplitude ${\cal G}_{ji}$ given by:
\begin{equation}
{\cal G}_{ji} = \langle j,t_j | i, t_i \rangle= \sum \exp [{ i S_{ji} /
\hbar }]
\label{Gij}\end{equation}
where the summation is over all paths from $(x_i, t_i)$ to  $(x_j,t_j)$
and $S_{ji}$ is the classical action evaluated along the path in question.

According to the postulates of quantum mechanics the propagator must
obey the group properties of completeness:
\begin{equation}
{\cal G}_{ji} = \sum_k {\cal G}_{jk} {\cal G}_{ki}  \ \ \ \ \ t_j \ge t_k
\ge t_i
\label{comp}\end{equation}
and unitarity:
\begin{equation}
{ \sum_k {\cal G}_{ki}^*{\cal G}_{kj} = \cases { \delta_{ij},
& if \ \ \ $t_i=t_j < t_k$ \cr
 {\cal G}_{ji}, & if \ \ \   $t_i < t_j < t_k$ \cr {\cal G}^*_{ij}, & if \
\ \ $ t_j < t_i < t_k$ \ . \cr}}
\label{unit}\end{equation}
Completeness asserts that if one examines ${\cal G}_{ik}$, then the
particle will have been at some position at any intermediate time $t_j$.
Unitarity is the statement that it is possible to reverse the time
evolution of a system so as to reconstruct an earlier state of the system
given the state at a later instant of time. Unitarity can be viewed as
being equivalent to conservation of probability.  It should be noted that
completeness and unitarity ensure that the time evolution of a system is
described by elements of a group, since the additional axiom of
associativity is clearly satisfied as a consequence of (\ref{comp}).  If
the Hamiltonian exists, and it is Hermitian, then completeness and
unitarity are trivially satisfied, as H is a generator of the Lie algebra
associated with the group of time evolution.

As an explicit example consider the propagator $K_{ji}$ of a free non
relativistic particle of mass $m$ propagating in a flat spacetime \cite{r15}.
\begin{equation}
K_{ji} = \cases { \left ( { m \over 2 \pi i \hbar (t_j-t_i) }
\right )^{{3\over 2}}
\exp \left ( { i m (x_j - x_i)^2 \over 2 \hbar (t_j-t_i) } \right ),
& if $t_j > t_i$ \cr
&     \cr
0, & if  $t_j < t_i$ \ . \cr}
\label{free}\end{equation}
$K_{ji}$ vanishes if $t_i > t_j$ since non relativistic particle propagates
only to the future. Since we are dealing with a non relativistic
propagator, the light-cone is the line $t=const$, that is a particle
located at $(x_i,t_i)$ can propagate to any point in which $t>t_i$.
Clearly $K_{ji}$ obeys (2) and (3), and can be derived by either
Hamiltonian methods or by path integrals.

Now we wish to study what happens quantum mechanically to particles
traveling in a spacetime that has closed timelike lines. Firstly, we need
to find a model spacetime in which calculations are straightforward but
nevertheless has the properties of a realistic time machine. We therefore
consider a flat spacetime in which we identify two selected spatial regions
$V_-$ and $V_+$ in such a way that for each point $(x_-, t_-)\in V_-$ we
match another point $(x_+, t_+) \in V_+$ with $t_+> t_-$.  This
identification mimics, in a simple way the effect of connecting two
timelike separated regions by a wormhole and hence, by convention we refer
to the connection between $(x_+, t_+)$ and $(x_-, t_-)$ as the ``wormhole."
The identification is so that a future-directed timelike line arriving at
$(x_+,t_+)$ emerges at $(x_-,t_-)$ again traveling toward the future and a
future-directed timelike line arriving at $(x_-,t_-)$ will emerge at
$(x_+,t_+)$.  We call this a two sided wormhole \cite{r16}. We denote by
$W_{-+}$ and $W_{+-}$ the propagators ``inside" the wormhole. Since we just
identify the points in $V_+$ and $V_-$, $W_{-+}$ and $W_{+-}$ degenerate to
an identity function.

Had there been no wormhole then the propagator in the time-machine
spacetime, ${\cal G}_{ji}$, would simply become $K_{ji}$ given by equation
(\ref{free}). However, it is not too hard to evaluate (\ref{Gij})
explicitly in the time-machine spacetime which we are considering, because
it is simple to find all possible paths by which a particle can propagate
from $(x_i,t_i)$ to $(x_j,t_j)$. We will explicitly calculate ${\cal
G}_{ji}$ for the case that $t_i < t_- < t_+ < t_j$. The possible paths are
then labeled by the number of times $n$ that the particle traverses the
wormhole, and the contribution to ${\cal G}_{ji}$ from all paths with fixed
$n$ is ${\cal G}_{ji}^{(n)}$.  For $n=0$ we have:
\begin{eqnarray}
{\cal G}^{(0)}_{ji}= K_{ji} - \int_+ d^3 x K_{j+} K_{+i} -
\int_- d^3 x K_{j-} K_{-i} \nonumber\\
+  \int_- d^3 x \int_+ d^3 x' K_{j+'} K_{+'-} K_{-i}
\label{G0}
\end{eqnarray}
where $K_{ji}$ is the ordinary propagator in the flat spacetime given by
(\ref{free}) and $\int_+$ ($\int_-$) denotes integration over the volume
$V_+$ ($V_-$).  The first term comes from all paths that go from
$(x_i,t_i)$ to $(x_j, t_j)$. However the second and third terms, which
represent the contribution of all paths from $(x_i,t_i)$ to $(x_j, t_j)$
via $V_+$, and all paths from $(x_i,t_i)$ to $(x_j, t_j)$ via $V_-$
respectively, must be subtracted off since any particle that arrives at
$(x_+.t_+)$ has traveled via the wormhole and emerged at $(x_-,t_-)$ (and
similarly for particles arriving at $(x_-,t_-)$).  In these subtractions we
have double counted the paths that in ordinary space would have gone from
$(x_i,t_i)$ via $V_-$ to $V_+$ and then to $(x_j,t_j)$ and those are added
in the last term.

For $n=1$ the calculation can be done in much the same way:
\begin{eqnarray}
{\cal G}^{(1)}_{ji}= \int_+ d^3 x \int_- d^3 x' \left [ K_{j-'}  -
\int_+ d^3 x'' K_{j+''} K_{+''-'} \right ]\nonumber\\
W_{-'+} K_{+i}  + \int_- d^3 x  \int_+ d^3 x' K_{j+'}  W_{+'-} K_{-i}
\nonumber\\ = K_{j-} W_{-+} K_{+i}  -  K_{j+'}  K_{+'-} W_{-+} K_{+i}
+ K_{j+} W_{+-} K_{-i}
\label{G1}\end{eqnarray}
where in the second equality we adopt the notation, like the summation
convention, that repeated {\it adjacent} indices are integrated over (thus
the indices behave like matrix indices).  The first term in (\ref{G1}) is
the contribution from a particle traveling through the wormhole once.  The
second represents paths that traverse the wormhole once and then would have
traveled to $(x_j,t_j)$ via $(x_+.t_+)$. Such paths are doomed to travel
through the wormhole once more and those will contribute to ${\cal
G}^{(n)}_{ji}$ with $n \ge 2 $.  The last term represents the contribution
of paths that reached $(x_-,t_-)$ and have traveled through the wormhole to
$(x_+.t_+)$ and from there to $(x_j,t_j)$.

Similarly one can construct the general ${\cal G}^{(n)}_{ji}$ for paths
traversing the wormhole $n$ times.  Unlike the previous two cases for $n
\ge 2$ we have contributions only from paths that began by reaching
$(x_+.t_+)$:
\begin{eqnarray}
{\cal G}^{(n)}_{ji}= K_{j-} W_{-+} \underbrace{ K_{+-'} W_{-'+}
K_{+'-''} W_{-''+''}....
K_{+''-'''} W_{-'''+'''}}_{{\rm (n-1) \ times}} K_{+'''i} \nonumber\\
- K_{j+} \underbrace{K_{+-'} W_{-'+} K_{+'-''} W_{-''+''}....
K_{+''-'''} W_{-'''+'''}}_{{\rm n \ times}} K_{+'''i}
\label{Gn} \end{eqnarray}

The complete propagator can now be evaluated in terms of $K_{ji}$ by:
\begin{eqnarray}
{\cal G}_{ji} = \sum_{n=0}^\infty {\cal G}^{(n)}_{ji} = K_{ji} +
(K_{j-} W_{-+} - K_{j+})(\delta_{++'} - K_{+-'} W_{-'+'})^{-1} K_{+i}
\nonumber\\ + (K_{j+} W_{-+}  - K_{j-} + K_{j+} K_{+-} ) K_{-i} \ .
\label{Gijf}\end{eqnarray}
Note that $\delta_{++'}$ has the dimension of $L^{-3}$.  $(\delta_{++'} -
K_{+-'} W_{-'+'})^{-1}$ is to be regarded as a matrix inverse.  The
propagator ${\cal G}_{ji}$ was derived assuming that $t_i < t_-$ and that
$t_j > t_+$. However, (\ref{Gijf}) holds for all possible time orderings
provided one recalls from (\ref{free}) that $K_{ji}=0$ if $t_j < t_i$, as
can readily be seem by considering all of possible time ordering of $t_i$,
$t_-$, $t_+$ and $t_j$.  The last term does not appear if we choose a one
sided wormhole, instead of the two sided one that we are considering
\cite{r16}.

To simplify the calculations we consider now $V_+$ and $V_-$ of spatial
extent $\Delta x \ll \hbar T / m X $ where $T$ and $X$ are any of the time
or length scales in the problem (e.g. $t_+-t_-$ ...).  Then, each of the
propagators can be taken to be constant over $V_+$ and $V_-$.  If $v$ is
the volume of $V_+$ (and $V_-$), then we find that \cite{r17}:
\begin{eqnarray}
{\cal G}_{ji}\approx
K_{ji} + v { (K_{j-}  - K_{j+}) K_{+i} \over 1 - v K_{+-} }
\nonumber\\ + v (K_{j+} - K_{j-}+ v K_{j+} K_{+-} ) K_{-i}
\end{eqnarray}
where we used the fact $W$ is an identity function and thus
$ K_{j \pm} W_{\pm \mp} \approx K_{j\mp}$.

To examine the completeness properties of ${\cal G}_{ij}$ we need to
evaluate $\sum_k {\cal G}_{jk} {\cal G}_{ki}$ for the various cases of
$t_j$, $t_k$ and $t_i$ greater or less than $t_-$ and $t_+$ and subject to
$t_i < t_k < t_j$. We use the completeness properties of $K_{ij}$,
(\ref{comp}), together with $\sum_k K_{+k} K_{k+} = 0$. The latter follows
form the fact that either $K_{+k}$ or $K_{k+}$ vanishes depending on
whether $t_k$ is greater then or less then $t_+$.  If $t_k>t_+$ or $t_k <
t_-$, completeness of ${\cal G}_{ij}$ follows directly from the
completeness of $K_{ij}$.  Hence, ${\cal G}_{ij}$ obeys the completeness
condition if the intermediate surface $(t=t_k)$ is chosen to be either to
the past or the future of the time machine. If however $t_- < t_k < t_+$
then
\begin{eqnarray}
\sum_k {\cal G}_{jk} {\cal G}_{ki}=K_{ji} + v \left ({K_{+-} K_{j+}
- K_{j-}  \over 1- v K_{+-} } \right ) K_{-i} \nonumber\\ + v  \left (
{K_{j-} (2 - v K_{+-}) - K_{j+}  \over (1-v K_{+-})^2 } \right ) K_{+i}
\end{eqnarray}
and completeness fails to be satisfied. This is because the particle can
cross such an intermediate surface exterior to the wormhole any number of
times.  This violation of completeness seems harmless as it happens only
while the time-machine is operating and completeness is recovered after the
time-machine has ceased to exist. Indeed had we considered a surface
intersecting the interior of the wormhole we would have discovered no
violation of completeness.

If we try to check unitarity either by using the unitarity of $K_{ji}$, or
by the explicit functional form of $K_{ji}$, (\ref{free}), we discover that
the unitarity condition is violated, unless all $t_i$, $t_j$ and $t_k$ are
to either the past or to the future of the time machine, in which case
${\cal G}_{ij}=K_{ij}$.  As an example of violation of unitarity we
consider the special case of: $t_i=t_j < t_- < t_+ < t_k$, i.e. the initial
and final points are to the past of the time machine and the intermediate
point is to the future of it (it will be too lengthy to write out here all
possible orderings of $t_i, t_j, t_k, t_-$ and $t_+$).  Then
\begin{eqnarray}
\sum_k {\cal G}^*_{kj} {\cal G}_{ki} = \delta_{ji} +
v \left ( {K_{+i} \over 1 - v K_{+-}} \right )
 \left ( -K_{+j} + K^*_{+j} + K_{-j}  -  K^*_{-j} +
v^2  K^{*\ 2}_{+-} K^*_{-j}  \right )
\nonumber\\ + v \left ( {K^*_{+j} \over 1-v K^*_{+-} } - K^*_{-j} \right )
\left ( - K^*_{+i} + K_{+i} - K_{-i} + K_{-i}^* + v^2 K^2_{+-} K_{-i}
\right )
\nonumber\\ +  v^2 K^*_{+-} K^*_{-j} \left [ v^2 K_{+-} K_{-i} +
\left ( 1- v K^*_{+-} \right ) K_{-i} + K^*_{+i} \right ] +
v^2 K_{+j} K_{+-} K_{-i}
\end{eqnarray}
and unitarity is broken \cite{r17}.  The physical reason for unitarity
violation is that the path integral for the reverse process is not given
simply by the complex conjugate of ${\cal G}_{ji}$ as is usually the case.
In fact it is not possible to construct an inverse to ${\cal G}_{ij}$ as
can be seen by attempting to construct the inverse of ${\cal G}_{ij}$ order
by order in $n$.

We have shown that in this simplified model for a time machine unitarity is
violated, although completeness is not. This means that there is no longer
a unitary time evolution operator for this system, and so the canonical
formulation of quantum mechanics is inapplicable.  However, it should be
noted that despite the fact that the group property of time evolution is
violated, the fact that completeness is preserved means that time evolution
can be described by a semigroup \cite{r19}.  This means that one can
predict the future from a given microscopic theory given data specified
before the time machine was formed, but it is impossible to reconstruct the
past on the basis of what one can describe to the future of a time machine.
This is the quantum analog of the results of Echeverria, Thorne and
Klinkhammer \cite{r13}.

{}From this simple picture, it is clear that quantum mechanics (as usually
formulated) breaks down in such spacetimes essentially because of the
existence of a closed timelike curves. If we try to extend our calculations
to more complicated or realistic cases- for example by having relativistic
particles- exactly the same phenomenon will occur because the pathologies
are associated with the non-existence of a Hamiltonian due to the
appearance of close timelike lines.  However, in these cases, it will be
more complicated to see explicitly how the difficulties arise.

The breakdown of quantum mechanics discusses here is very reminiscent of
the phenomenon of black hole evaporation \cite{r11}. If black holes
evaporate completely, then it seems likely that information about the
collapsing matter is annihilated, as a consequence of lack of a unitary
time evolution which manifest itself as a pure state developing into a
mixed state.

We would like to conclude with some speculations. Standard Hamiltonian
quantum mechanics is violated by closed timelike curves, so we can suppose
that if the laws of nature are truly quantum mechanical then it will be
impossible to construct time machines. If however, as has been suggested on
the basis of black hole physics, quantum mechanics breaks down when
gravitation is taken into account \cite{r20} (i.e. black hole evaporation
when non-unitary evolution of similar nature also appears to take place)
then we see no reason why it should not be possible to construct such
machines \cite{r21}.

\acknowledgments

It is a pleasure to acknowledge S. Coleman, J. Hartle, A. Strominger and K.
S. Thorne for enlightening conversations. We thank the Aspen Center for
Physics for hospitality while this research was done.  This research was
partially supported by a Center for Astrophysics fellowship (DSG), and
by the  Royal Society and Trinity College (MJP).


\begin{references}
\bibitem{r16} Note that ``inside" this (two-way) wormhole, in which particles
travel in both directions, time orientability is lost.  This can be avoided
by considering twin wormholes connections the regions $V_+$ and $V_-$ or by
choosing a one sided wormhole, i.e. one in which a particle that is
traveling towards the future that arrives at $(x_+.t_+)$ is transported to
$(x_-, t_-)$ and placed back in the spacetime there, but a particle that
reaches $(x_-,t_-)$ is not transported inside the wormhole to $(x_+.t_+)$.
Our main result holds also for the one-way wormhole.
\bibitem{r19} A semigroup , $S$, obeys all the axioms of a
group except that elements of $S$ no longer are required to have an
inverse.
\end{references}
\end{document}